**IT Risk Assessment for Group6 Healthcare Clinic- Report 2017**

| | | |
|---|---|---|
| STUDENT NAME: | **Abdi Karim Hilowle** | |
| STUDENT NAME: | **Israel Cole** | |
| STUDENT NAME: | **Jaye Jallow** | |
| STUDENT NAME: | **Liswani Sibamba** | |

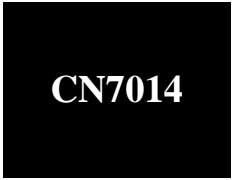

**CN7014**

**TABLE OF CONTENTS**:
   Abstract

ABSTRACT:

Information security and privacy in the healthcare sector is an issue of growing importance. The adoption of digital patient records, increased regulation, provider consolidation and the increasing need for information exchange between patients, providers and payers, all point towards the need for better information security. We critically survey the literature on information security and privacy in healthcare, published in information systems journals as well as many other related disciplines including health informatics, public health, law, medicine, and the trade press and industry reports.

1. INTRODUCTION:

The Group6 Healthcare Clinic carried out a task to conduct an Information Security Management System (ISMS) risk assessment base on ISO27001 standard. This risk assessment report is focus on a Healthcare clinic based in Newnham; it provides health services to the local community and surrounding areas. The Clinic runs daily clinics and it is a favourably busy public service organization. It has an IT infrastructure in place, which helps in the delivery of Information and communication services for staff in various departments. The clinic handles a lot of private information about its patients, statistic as well as sensitive service information, which needs to be safeguard against unauthorized access.

The healthcare service is being in operative within the United Kingdom providing services for both in and out patients that mainly depend on IT technology to perform their everyday tasks such as, patient's bookings containing sensitive data and information i.e. Date of birth, email address, bank card details, medical diagnostics results, home address and more.

The company's policy is already set out initially but with continuous threats and vulnerability there's concern to assess the overall management system in line with ISMS strategic plan model (PDCA Model), Confidentiality, Integrity and Availability of Patients records and data.

Therefore, it was vital on our findings to assess the risks through the recommended model to Investigate, identify the potential risks, plan of exercises and techniques to be carried out and what method to adopt in this case study after exploration of the threats, vulnerabilities within the organization structure. First is to find the critical success factor of risk, then map the identify values of risk are understood by risk analysis. Then next is to look at the probabilities, which is, what is the threat to the asset, who is the owner, what is the risk register saying of its value. This then leads to the vulnerabilities of the concerned threat and its likelihood to its risk values. Finally, we look at what Mitigation we must what we need – this leads to the control factor. Mitigation gives us the cost factor – the cost at which a control are put in place like a firewall, an extra monitoring engineer, staff training, reputation management, damage control. The next is to understand security risk on its own merits/demerits and benchmark to know values acceptable in the industry sector that you work with like NHS, Patient Care and many others. After the identification of these risks we will further determine what appropriate measures should be adopted and recommended solutions that will be implemented for robust security framework in compliance to data usability, scalability, redundant and scalability ensuring better business continuity on ISO27001 standard.

## 1.1 Case study Overview including organizational Diagram:

The Group 6 Healthcare Clinic is located in Forest-gate under the Newnham borough. The clinic operates under the guideline of the NHS and it has a radiology scanning specialized Centre for patients. It is a highly computerized clinic with high dependence on its computer infrastructure for its day-to-day running and keeps all its records electronically. It has a complete workforce starting from high-level management to casuals and temporary workers. The organization has various departments with different specialties; among them is the radiology department, which runs bespoke software for scans, X-rays and other radiology related functionalities. An in-house team manages the software, which is under the jurisdiction of the IT department. The radiology software is a critical part of the department as well as the entire clinic, any malfunction would result in cancelled appointments, endangering of lives and above all, a massive backlog of patients on the waiting list.

Figure 1 below is the organogram structure for Group6 Healthcare Clinic organization.

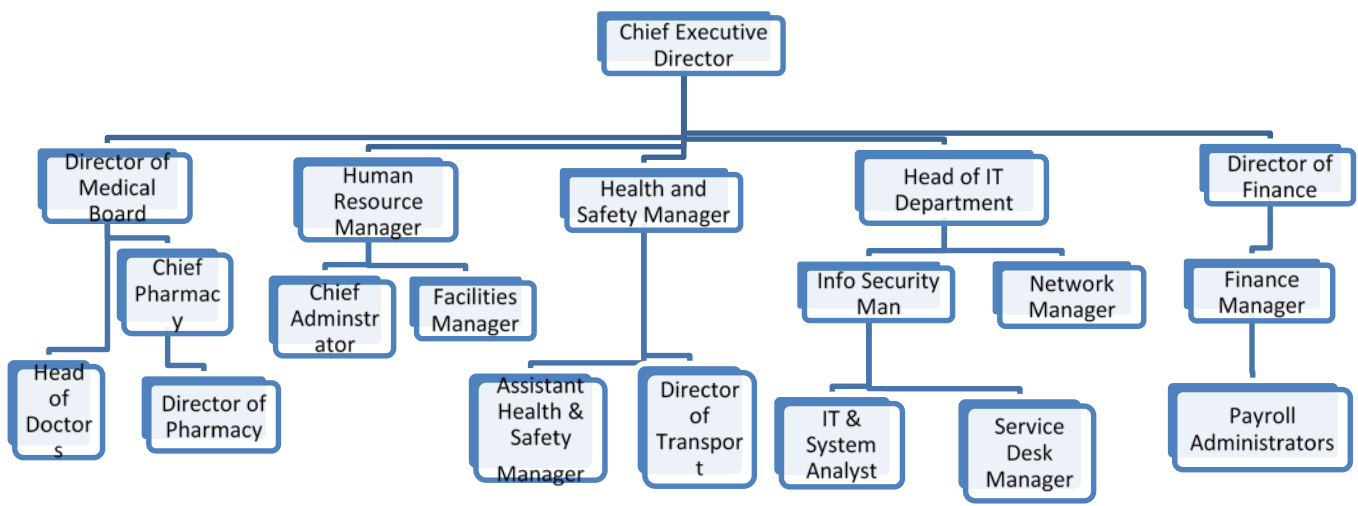

From the diagram above the case study team have Identify five (5) different assets to the clinic, these assets are: The Staff/Temps/Volunteers, Patient Records, GP6 Radiology Software, Servers (Physical) and Company Reputation. These assets categorically have Asset owners and Asset values. It is important for the company's assets to be distribute amongst employers, if necessary to help reduce the risk of a weak system, according to (Shon Harris, Fernando Maymi 2015), that a weakness is a system that allows a threat source to compromise its security, this can be a software, hardware, procedural, or human weakness that can be exploited.

## 1.2 Motivation and Justification:

Today's world needs more dynamic risk assessment to base line methods. By setting up risk register, which is done and then mapped to the owner of the asset like if its server it is the IT infrastructure team, desktop the IT support team, Mobile device is the Mobile device management Team and so on. Next controls are place on this team to the assets life management cycle. Training is reinforce to let the owner or user know that it is mandatory, obligatory to follow the procedure due to the risk identified and the mitigation methods laid down.

The risk management team of the NHS – CRO, CEO, Network Manager, other executive look at the risk register and then sorts the risk on the overall rating calculated and makes a priority of how to mitigate the risk while keeping cost low. The task is then assigned to the asset owner and the method is follow until any risk that is having 50% is tackle. The rest of the risk is then handle by a mission critical basis, balanced with actions, and assigned to the owner of the asset. This is then followed up my meeting to see progress of implementation and if they likelihood has gone down like firewall in place and zoning done threats have been reduced, VLAN split has decrease the spread surface of the attack.

The next most important fact is to understand the cost value of the IT Security - risk management. There is no known good model of costing to justify the right way of investment and its tangent value of Returns of Investment. It is good for healthcare to go on the internet but the impact of data breach etc. The internet model might bring down the cost of operation and bring in better service but will expose the organization to various layers of cyber security threats, which need to be, mitigate. The effective value are not outline – a case study on this effectiveness is being reflect in Economic valuation for information security investment (Schatz et al 2016). This article gives a benchmark to various article, website, and publication for benchmarking the process.

## 1.3 Why Risk Assessment is Important (Impact):

Risk management begins with mapping business objective to risk objective - they is a directly linked. This analysis leads to the risk appetite being roll off where risk is identifying by the organization. This risk is then map on to the risk register and assigned owners with possible threats and vulnerabilities. Like Human Resource contributes to various risk factors in a health care organisation. Most of the Health care organization depends on the KEY-Risk-Of-Specialty-Dr. Their illness, death, social issues can poise a major threat to the Healthcare organization. The Chief of Medical Facilities identifies this as a risk to be monitor also to be balance. As having more than one heart, eye, and orthopedic surgeon on board takes care of this issue, also they can have a second opinion on board for no cost. This process of identifying the risk and mapping it on to the risk register is from *Risk appetite framework* to management of risk function.

To build an organization one must have a wider understanding of the language of risk, a risk committee is can be form, and are led by Chief Risk Officer (CRO) and other board members. The primary task of this CRO is to make everyone understand how to talk the language of risk. This involves the overall risk management – first is the *tools used like Risk Register,* any software, frameworks, etc. To bring good quality the Risk culture should be develop where in appropriate training implement to the completely corporate how to deal with the issue.

The risk should always begin with the risk *appetite framework* and the overall governance of the organization toward risk is to be align to a critical success factor.

The various risk drivers identified by the risk appetite framework are then map to a *risk profile*. This is what regulators look at. The risk profile follows various industry standards of calibration based on likelihood and asset value to define tolerance factors of limits and triggers to handle the situation.

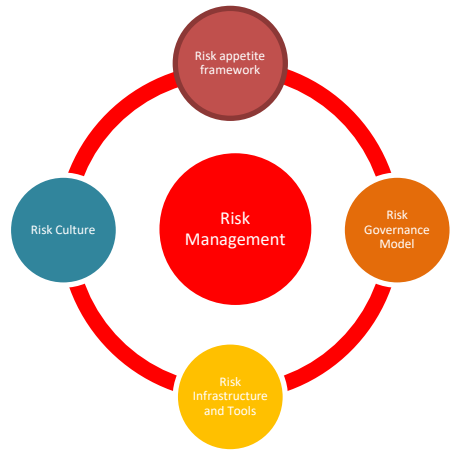

**Figure 2 - Risk Management Model for healthcare system:**

For the current healthcare organization, based on the above model - we have developed a Risk register and populated, with known risk values is outline below. The major owners in the risk management team are HR, Chief Administrator, IT and System Analyst Manager, Network Manager, CEO. Various threats and vulnerabilities have been identify for these owners as part of Risk Appetite Framework.

Human resource is a major area of concern we have identified three type of threats. They can pose problems to the business activities and for each threat carry three vulnerabilities. The overall classification of this is value=2 has this risk is more at an individual level and not having a tenor of more than 100k.

Medical Department is the most critical area of concern and it can influence the business objective very directly and spoil the reputation of the healthcare, a value of (5) which is the highest. The risk is overall of the company and can lead to patient death resulting in above 100k values claims.

IT and System Analyst Manager have a key role in protecting data especially with UK being a part of Europe Union the health care needs to follow the GDPR of the EU. Compromised information can lead to charges and penalties for a healthcare leading to cost of (500k-1000k) that is why this is a major risk and given the value (4). The threats are illegal use, software failures, virus attack with their likelihood and vulnerabilities

Network Manager is very important in day-today world of risk – they take care of most risk in the environment. Virus, Trojans, hackers, spam, dos, DDos. The risk is handle at various levels within the security layers of open system interface (OSI) from the ISO. Various logs like syslog, centralized software, netflow, Cisco IPS, Fortinet, Checkpoint software, quantify the risk. This is then map to various co-relation attack. Like virus from one pc spreading to another in absence of VLAN, firewall zoning, sub-netting and virtualization, cloud, etc. The network managers maintain their own standards of matching threats and vulnerabilities and report to the RISK CRO for quantified action and its value like having a robust Next Generation Firewall (NGFW) against the number of attacks.

CEO keeps a constant tab over the source of risk to any organization and monitors risk assessment strategy. The CEO is responsible for keeping the status quo – reputation of the healthcare.

The organization needs to work like a team. The threat vector need to be map to the number of employees – quantification, type of employee related vulnerabilities and their likelihood – which is a quality matter. THE CEO/CRO need to next look at the total cost of the risk which is the cost of mitigating the risk – one time cost, remediate the issue every time the damage is done multiplied by the number of times it occurs in a year. This overall gives us a tangent value to make a close to accurate decision to the board for approval.
In a NHS environment, the data is the most important information we have. Human error, theft are the main viewpoint of risk alongside a few other factor like hackers attacking the system.

In addition, Human error can be tackle by training the staff or having induction training by outline within the company's policy etc. Theft is taken care by implementing various policies like data leak policy DLP on all the software. Software of the NHS is another major area of concern, which is prone to virus attack, software upgrade related issue, documentation, contingency, backup, restore, release management, User acceptance testing before rollout. This all involves a good cost to baseline the software to standard defined in the industry. Server, Network Equipment is the next main factors, which with VMware, Hyper-V has moved into cloud computing model changing the dimension of the risk assessment to management and strategy. Reputation of the organization to prevent data leak, maintain standards, other regulation is also of very important consideration to the risk assessment strategy. This leads us to Mitigations (Existing, Being Enhanced, or Potential Future).

## 2. Scope of the Risk Assessment in Compliance with ISO27001:

As illustrated in the organization's management structure, The Group6 healthcare Clinic has various departments that help in ensuring the smooth delivery of service to its patients and the members of the public. In this case, study, we have decided to narrow down our security management scope to the IT department. The IT department responsible for the delivery of information and telecommunication technology for the clinic, which includes supporting the IT infrastructure i.e. network support, desktop support, intranet and internet.

Information security is a fast-growing discipline. The protection of information is of vital importance to organizations' and governments, and the development of measures to counter illegal access to information is an area that receives increasing attention. Organizations and governments have a vest interest in securing sensitive information and hence in securing the trust of clients and citizens. Technology on its own is not a sufficient safeguard against information theft; staff members are often the weak link in an information security system. Staff members can be influence to divulge sensitive information, which subsequently allows unauthorized individual's access to protected s y s t e m s .

## 3. Identification of the Asset and their Asset owner:

According to (Calder and Watkins, 2015), information is vital in any organization in the modern business and are prone to various types of threats from hackers, online fraud and virus attacks. Therefore an insider threat for example can be an umbrella for a number of different malicious acts carried out against that organization. This covers a gamut of actions, from disgruntle employees in deleting sensitive records or information to even state sponsored espionage carried out via someone close to the organization, and everything in between.
Insider threats are much more heavily cloak than external threats by the organization suffering them. This could be due to the very sensitive nature of an internally initiated breach. Because of this, companies avoid the use of the law, preferring to deal with the issue, differently.

**The case study group identifies 5 assets as requested by the organization:**

**Asset 1:    Staff/Temps/Volunteers:**
Staff members are a significant part of any organization and department and are classify as part of the assets, and loss or unavailability of staff is a risk to the organization. The various factors that would lead to loss or unavailability of staff; these have been listed down as vulnerabilities in the risk register.

Security in an organization is people problem, therefore staff morale and awareness are of importance ensuring that employees has the relevant information that should be enforced by managers to assist in the compliance of the organization policy recommended by regulatory bodies such as GDPR or likewise.

**Asset 2:  Patient Records:**
The policy aims at checking health information security on patient records of the clinic and department to follow approved standard ISO27001. To achieve that security measures are identified and make recommendation to improve        the quality of healthcare by making it more personalized and reducing costs and medical errors ensuring security and privacy to the systems are not compromised but socially acceptable.

**Asset 3: GP6 Radiology Software (In- house):**

The software application is develop in house to meet with patient and staff needs by the IT department to facilitate the effectiveness of the company demand.

 The clinic uses bespoke software in the radiology department, it is an important aspect of the services provided by the clinic, and it is a closely monitored asset of the system, with access only granted to staff users after a short training. Any disruption to the operations of this software would result in patients having to wait long hours or days before they could have their urgent or routine radiology scans. This may attract unnecessary attention from stakeholders, bad reviews by patients, may also filter through to local media houses.

**Asset 4 – Servers (Physical):**

Servers provide a variety of internal and external user services in organizations, Companies are at risk of cyber-attack from hackers, viruses and online fraudsters for Information or data which are constantly vulnerable to security threats, due to data sensitivity.
Secure servers help organizations and businesses conduct secure and private network transactions. Until recently, e-commerce opportunities were often lost because of online user security concerns. However, the growth of online retailing has expanded requirements for security and measures geared toward preventing malicious attacks (like phishing and hacking).

**Asset 5: Company's Image and Reputation:**

Every organization prides itself for its well-established reputation; it is very important that the reputation of the clinic. Any damage to this reputation would lead to the loss of confidence by the patient and the member of the public; this would also bring about bad publicity in the media.

**1. Staff/Temps/Volunteers:**
**Threats:**
I. Illness affects most organizations in the UK and across the globe. We have scaled down our most susceptible vulnerabilities that are likely to affect our staff and the smooth running of our organization.

II. Death is can happen at any time, organizations are usually affected if a staff member is lost by death,

III. Social Engineering: Social engineering attacks involve some form of psychological manipulation, fooling unsuspecting users or employees into handing over confidential or sensitive data. The most common form of social engineering involves email or other communication that invokes urgency, or fear in the victim, leading them to promptly reveal sensitive information, or click a malicious link, or open a malicious file. Because social engineering involves a human element, preventing these attacks can be tricky for organizations. Below are some of the vulnerabilities associated with social engineering; Financial Incentives, Trust and Social media activities:

**Vulnerability:**
**Illness:**
I. Super Bug: is a very serious illness epidemic. If a staff is attacked by the disease and a competent staff for that matter, is a setback for any organization, much more the clinic in particular.

II. Flu Virus: is common and affects a lot staff in the UK. It a highly classified epidemic and can spreads so quickly and staff can come in contract the virus whether the seasonal winter flu or the regular cold, this can have attributed to the working environment by coming in contact with patients or fellow staff with the virus.

III. Exposure to X-ray:
This may affect front-line staff working in the radiology department over a long period, this is less likely risk but a risk worth put into prospective for the general well-being of staff.

**Death**
I. Drug overdose: According the Office of national statistic, 6,803 deaths were drug or substance abuse related between 2014 and 2016 in England alone. It is a serious concern that could lead to death and if it happens to staff, then we consider it as not-recoverable asset. The use of illegal drugs or not-prescribe drugs is a consequence and can lead to death. If a clinic staff undergone this lifestyle, in most cases is course by stress or domestic violence from home.

II. Trips and falls happen in our day-to-day life, it could be at work or at home and some falls result in critical injuries that eventually lead to death. In a circumstance where a staff trip and fall got injured that resulted in death relates to health and safety issue this may considered that the management lacks some policy discipline in their duties weather they are enforcing the company policy. As a result, this can be very costly and hence causing a serious breach.

III. Traffic Accident (RTA): Traffic accidents are a common occurrence on British roads; hence, this has been included as one of the vulnerabilities, in the event that a member of staff is involved in a fetal road traffic accident, It is referred to as acceptance of risk control strategy as the clinic will do nothing, but accept the damage.

## 2. Reputation:

**Threat: Patient Records**

I. Financial incentive: As we all know, that protecting assets is important. The main threat to these assets are the physical security, if they are compromised which will interruption service. If staff's demand considered, they will compromise with patient's record and a sign of negative reputation. If the clinic previously agreed to increase financial incentive, may be due to budgetary aspect they could not dwell on their promises,

II. Revenge: Record theft are common now a day if management fails to promote competent staff or a staff is harassed, and they hold a key position within the clinic, they can revenge by comprising patient's records without delay.

III. III .Data Diddling: Another very widely used data theft in today's computing is data diddling, this is where the attacker gains access to the system and make changes in the data either during by storage, input, output or transaction. They attacker also will use this occasion to compromise the data, if the clinic did not secure their system, we expect that data diddling will take place at any given time and is a serious vulnerability.

## 3. GP6 Radiology Software (In- house):

**Threat:**
**Illegal Use:**

I. Malicious code: If an intruder or unauthorized person gained access to the clinic's network system, they will inject malicious code to interrupt the system. In most cases, they will take over the entire system and it is usually hard to recover the records if backup where not created.

II. Piracy: Because clinic's radiology software is an in-house built-in. Security is one aspect to be considered; this is where a proper protection be recommend avoiding a disgruntled staff access the system and compromising it software and sell it to other company.

III. Intruder: unauthorized person who gain access to a system, either by hacking or using tricks to course denial of service to the authorize. Intruders are every day looking for systems to access, the clinic must secure entire system, if these measures are not instituted, intruded can easily gain access to the system.

**Software Failure:**

I. Software Testing: whenever software is developed and not tested, it will fail at one point or forever. The Clinic's Radiology software is in-house designed and developed during the process it should be tested, if not consider it crash to the core. This is a vulnerability which is costly to the clinic if has occurred.

II. Pseudo Flaws: is a method of hacking is common into today's glob of computing. It is where an intruder, will implant a loopholes of honeypot system, by emulating a well none operating system, intentionally in pretending as a legitimate user. The clinic IT team must make sure that their system is secure to protect such activities to occurred, if they allowed it to occur, it will be costly and a negative reputation to management as patient's record will be lost of hacked.

**Virus Attacks:**
8. Susceptibility: This is a form of virus, which appears in a link form injected mostly into application software, appearing genuine to open, but a virus will pop-out to destroy the system. The clinic must make sure that such links must not be compromise and advice all staff and then continue to monitor the activities within the system while in operation. If it is not taking care off, can course a serious cost to the clinic.

9. Masquerading attack: this is an old technic and it is still in use but update. inciting (Stewart, et al, 2015) that in an offline world, a teenager will borrow a driving license from an old sibling to purcha.se alcohol, this are the same way in the computer security, where an attacker will borrow an identity of a legitimate users and systems the trust of a third party. The clinic must put in place employ proper security to their system to overcome this threats and vulnerabilities.

10. IP Spoofing:
This is where an attacker configures their system to a trusted system IP Address in order to gain access to a third-party system. If a network does not implement a better filtering system to monitor their network system, any time an attacker can easily gain access to change the configuration and do damage to the network. A network administrator must configure a filter to every network perimeter to avoid such vulnerability.

**4. Servers (Physical):**
**Threat**: Fire

**Vulnerability:**
I. Inadequate Fire detector: Since a Server are the backbone of every network system, it acts as the mother of communications, and they must be secure to an extent for unauthorized users. We also contemplated that there are certain threats that can affect the server premises, fire can be a serious consequence to the server, if not properly instituted, the lack of inadequate fire detector can be consider serious vulnerability to the clinic.

II. Electrical Failure: In today's computing, an uninterrupted power supply units (UPS) are available in case of electrical failure, if the clinic has no means of putting this equipment in place relaying on national electricity, always expect the unexpected, it can occur, whereby they entire system will be down and may be recoverable, it will be costly if not consider.

III. Extreme temperature: this is a serious vulnerability to the server room and system, if it is not considering it will be disastrous to the entire clinic because it will course fire and will destroy the entire network system.

## 5. Patient Records:
   **Threat:** Human Error, DOS Attacks

Vulnerability:
Human Error:
  I. Distraction: During patient's records processing period (input), distraction by human factor can course typing-error. It is always advisable that those involve in data input should be confined to a special or noise-proof office and restrict access to avoid distraction. If not considered there is an expectation of wrong output to important documents.

  II. Data leakage: this are normally carried out by an insider, mostly disgruntled staff man easily compromises with patient's record may be for personal gain or if their expectation not met by the management team, this could be a promotion overdue or allocated to someone more junior, these are all possibilities of data-leakage. It is a serious vulnerability that the clinic should always consider and prepare to motivate the staff.

  III. Workload: organization should follow the best practice of enrolling enough staff and competency respectively, this is all factors that management should consider and must be address and monitor in day-to-day running of the organization. Records input are an ideal aspect of every responsible clinic must be very seriously and create and enclose and protect it, as it is the back-born of the clinic.

DOS Attack:
  I. SYNC Attack: Distributed denial-of-service (DDoS) this is common in today is computing and it is also identical to Dos attack, the difference is that it has a greater volume. The attacker will employ (SYN, ICMP and DNS) to hijacked or zombie computers, it vulnerability are occasional attack and by sending malware, if the user clicks a link, your system is redirected, and the attacker will build their own botnets and your system then is taken over. Where patient's record is processed or stored, there should be a better control mechanism to protect the equipment and to avoid intruders to access the system.

  II. NTP: Network Time Protocol (NTP) is the oldest network protocol used on the internet, but still in use in today's computing. This works in such a way that log files are stored locally on the corresponding device, it is very easy for an attacker to alter the log files of whatever devices they compromise. The only way to overcome this issue or project is to centralize the location of the log files across the clinic to minimize the risk. Having not distributed the NTP, it can be disastrous and can be costly.

  III. Botnets: is a way of infecting a computer system by injecting malicious code or malware into the keyloggers to capture the user's keystrokes. If that is achieved, then the entire server is taken-over and is remote control and a browser based activities. The only way for the clinic to prevent this actives is to install anti-malware antivirus and built a strong security mechanism and they must keep system up to date on patches. If the clinic follow this security measures, they will be safe or reduce the damage if it arise. Failure to adopt the check and balance render their servers insecure and costly that can even result to data lost.

**Recommendations of Risk Assessment over Baseline:**

A baseline approach are mainly use by small organizations, is can be consider as industry's best practice, which can be cheap to implement and is replicable. On the down side, it gives no special consideration to organizations and provides too much or too little security but it's also know to implement safeguards against most common possible threat. The baseline approach is not appropriate and it is best to start with risk assessment.

Like for example the baseline approach is when there was no mobile device connected to the organization corporate network and now we have them connected and we have two classifications of corporate device. This new development has failed the baseline approach so we need to do risk assessment again - scope-it, risk factors, occurrences, cost. The base line approach is easy and a straight jump to implementation of risk mapping but you might miss something like the security of mobile device in the scoping. Baseline models like software BART (Baseline-Risk-Assessment-Tool) have been define very successfully by various companies like Microsoft, Oracle, Cisco, etc. but only cater to the boxes or software they supply. For complete organization security, there are various new BART software in the market. The risk assessment module needs some additional resource to define the baseline model to be implemented – this people should understand system design, risk management, risk tree, Mitigation. Cost factor (Risk, Mitigation).

Risk assessment is the most important part of any system. It has a fundamental basic wherein the whole system is looked into – Threats are identified (like attack on server in DMZ), based on the threats vulnerabilities are identified (Ports on the server which is Vulnerable, DOS, DDOS, SYN Attack), Mitigation is have a firewall like cisco ASA, Cisco NGFW, Fortinet, Bluecoat. The risk assessment is a complete wholesome view by an expert who understands the whole system or studies the whole system. The model he develops can be baselined and are in similar environment, which is static and not dynamic. So if the variables changes of the scope, threats, vulnerabilities, mitigation, etc. the risk assessment should be redone by a security expert who then baselines it. The baseline has develop overnight – they are a time series tested documented model – like CMMI. Hence, Risk assessment by an expert is a good measure.

**Justifying a Qualitative and Quantitative Approach towards Risk Assessment:**

The case study adopts two-method approach, qualitative and quantitative to focus on the subjective to evaluate the likelihood of an incident to occur or the likely event that this incident is eminent to take place. According to (Calder and Watkins, 2012) in any likelihood event when it is necessary to use numerical strategy connecting figures that is associated with risk to ascertain any potential cost(money) and values relating when an event happens. It is viable that such estimate will achieve a great result depending on cost analysis recommended.

Therefore, it has a pitfall depending on the measurement of this numerical figure for instance; in money, related condition where information is more promptly measured quantitative effect may be vague.

Nevertheless, an advantage of qualitative risk assessment approach will determine on how to prioritize any given risks by identifying the potential that need immediate actions and addressing possible threats and vulnerabilities. On the other hand, there are disadvantages when adopting a qualitative approach. It is that it does not provide specific, quantifiable measurements of the magnitude of the impacts, therefore difficult to address the cost –benefit assessment of any recommended controls. On the other hand, the approach lacks empirical data, and the difficulty to put metrics on the data available is the reason behind organization is using this approach.

In comparing these two methods of approach, our case study recommend that we adopt the semi qualitative and quantitative (hybrid) approaches within the risk assessment process the Group 6 Healthcare Clinic. To enable the company using estimates of asset values and assigns predefined risk to a quantitative scale that in turn can adopt the qualitative labels as mentioned by (Whitman & Mattord, 2010, p.325). This will be the case study group to deal with a better and achievable result for the said company.

The diagram below justifies hybrid approach of Qualitative and Quantitative adopted by Group 6 Healthcare Case Study Team.

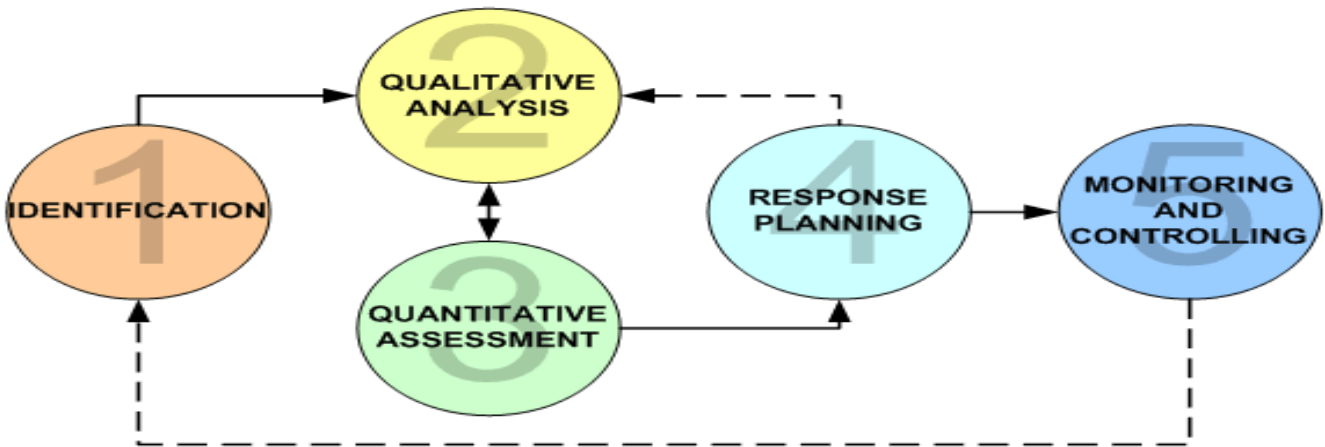

**Stages of Populating Risk Register and their Assets:**

## Stages of Populating Risk Register and their Assets:
## Risk Register Table

| Asset No | Asset Name | Asset Owner | Asset value | Threats | Likelihood | Vulnerability | Likelihood | Risk | Existing Control | Net Risk | Certainty | Overall |
|---|---|---|---|---|---|---|---|---|---|---|---|---|
| 2 | Patient Records | Chief Administrator | 5 | Human Factor | 5 | Distraction | 10 | 250 | 65% | 87.5 | 90% | 96.25 |
| 2 | Patient Records | Chief Administrator | 5 | DDOS Attack | 8 | Bootle neck | 6 | 240 | 65% | 84 | 90% | 92.4 |
| 2 | Patient Records | Chief Administrator | 5 | DDOS Attack | 8 | Weak firewall Config | 6 | 240 | 65% | 84 | 90% | 92.4 |
| 2 | Patient Records | Chief Administrator | 5 | Human Factor | 5 | Work Load | 9 | 225 | 65% | 78.75 | 90% | 86.625 |
| 3 | Gp6 Radiology Software (In-house) | IT & System Analyst Manager | 4 | Virus attacks | 6 | Susceptibility | 10 | 240 | 70% | 72 | 85% | 82.8 |
| 2 | Patient Records | Chief Administrator | 5 | Human Factor | 5 | Data leakage | 8 | 200 | 65% | 70 | 90% | 77 |
| 1 | Staff/Temps | HR Manager | 2 | Illness | 8 | Flu virus | 10 | 160 | 60% | 64 | 80% | 76.8 |
| 4 | Servers(Physical) | Network Manager | 4 | Unauthorise Access | 6 | Unsecured server room | 7 | 168 | 65% | 58.8 | 83% | 68.796 |
| 2 | Patient Records | Chief Administrator | 5 | DDOS Attack | 8 | Sync Attacks | 4 | 160 | 65% | 56 | 90% | 61.6 |
| 1 | Staff/Temps | HR Manager | 2 | Illness | 8 | Super Bug | 8 | 128 | 60% | 51.2 | 80% | 61.44 |
| 2 | Patient Records | Chief Administrator | 5 | Theft | 6 | Financial incentive | 5 | 150 | 65% | 52.5 | 90% | 57.75 |
| 2 | Patient Records | Chief Administrator | 5 | Theft | 6 | Revenge | 5 | 150 | 65% | 52.5 | 90% | 57.75 |
| 2 | Patient Records | Chief Administrator | 5 | Theft | 6 | Data Diddling | 5 | 150 | 65% | 52.5 | 90% | 57.75 |
| 3 | Gp6 Radiology Software (In-house) | IT & System Analyst Manager | 4 | Virus attacks | 6 | Masquerading of User ID | 6 | 144 | 70% | 43.2 | 85% | 49.68 |
| 3 | Gp6 Radiology Software (In-house) | IT & System Analyst Manager | 4 | Virus attacks | 6 | IP Spoofing | 6 | 144 | 70% | 43.2 | 85% | 49.68 |
| 5 | Reputation | CEO | 5 | Patient records | 6 | Theft | 6 | 180 | 75% | 45 | 90% | 49.5 |
| 4 | Servers(Physical) | Network Manager | 4 | Unauthorise Access | 6 | Policy | 5 | 120 | 65% | 42 | 83% | 49.14 |
| 4 | Servers(Physical) | Network Manager | 4 | Unauthorise Access | 6 | Registry log | 5 | 120 | 65% | 42 | 83% | 49.14 |
| 3 | Gp6 Radiology Software (In-house) | IT & System Analyst Manager | 4 | Illegal Use | 5 | Malicious code | 7 | 140 | 70% | 42 | 85% | 48.3 |
| 3 | Gp6 Radiology Software (In-house) | IT & System Analyst Manager | 4 | Illegal Use | 5 | Piracy | 7 | 140 | 70% | 42 | 85% | 48.3 |
| 3 | Gp6 Radiology Software (In-house) | IT & System Analyst Manager | 4 | Illegal Use | 5 | Intruder | 7 | 140 | 70% | 42 | 85% | 48.3 |
| 5 | Reputation | CEO | 5 | Patient records | 6 | staff sabotage | 5 | 150 | 75% | 37.5 | 90% | 41.25 |
| 5 | Reputation | CEO | 5 | Patient records | 6 | data leakage | 5 | 150 | 75% | 37.5 | 90% | 41.25 |
| 5 | Reputation | CEO | 5 | Staff | 5 | industrial action | 6 | 150 | 75% | 37.5 | 90% | 41.25 |
| 5 | Reputation | CEO | 5 | Staff | 5 | Incompetent Staff | 6 | 150 | 75% | 37.5 | 90% | 41.25 |
| 4 | Servers(Physical) | Network Manager | 4 | Temperature | 5 | Poor ventilation | 5 | 100 | 65% | 35 | 83% | 40.95 |
| 4 | Servers(Physical) | Network Manager | 4 | Temperature | 5 | inadequate temperature control | 5 | 100 | 65% | 35 | 83% | 40.95 |
| 4 | Servers(Physical) | Network Manager | 4 | Temperature | 5 | poor designing | 5 | 100 | 65% | 35 | 83% | 40.95 |
| 3 | Gp6 Radiology Software (In-house) | IT & System Analyst Manager | 4 | Software Failure | 5 | software Testing | 5 | 100 | 70% | 30 | 85% | 34.5 |
| 3 | Gp6 Radiology Software (In-house) | IT & System Analyst Manager | 4 | Software Failure | 5 | Pseudo Flaws | 5 | 100 | 70% | 30 | 85% | 34.5 |
| 3 | Gp6 Radiology Software (In-house) | IT & System Analyst Manager | 4 | Software Failure | 5 | Maintenance | 5 | 100 | 70% | 30 | 85% | 34.5 |
| 5 | Reputation | CEO | 5 | Staff | 5 | Repudation | 5 | 125 | 75% | 31.25 | 90% | 34.375 |
| 1 | Staff/Temps | HR Manager | 2 | Social Engineering | 5 | Trust | 7 | 70 | 60% | 28 | 80% | 33.6 |
| 1 | Staff/Temps | HR Manager | 2 | Illness | 8 | Exposure to X-ray | 4 | 64 | 60% | 25.6 | 80% | 30.72 |
| 4 | Servers(Physical) | Network Manager | 4 | Fire | 3 | Extreme temperature | 6 | 72 | 65% | 25.2 | 83% | 29.484 |
| 1 | Staff/Temps | HR Manager | 2 | Social Engineering | 5 | Social Media Activities | 6 | 60 | 60% | 24 | 80% | 28.8 |
| 4 | Servers(Physical) | Network Manager | 4 | Fire | 3 | Inadequate Fire detector | 5 | 60 | 65% | 21 | 83% | 24.57 |
| 4 | Servers(Physical) | Network Manager | 4 | Fire | 3 | Electrical Failure | 5 | 60 | 65% | 21 | 83% | 24.57 |
| 1 | Staff/Temps | HR Manager | 2 | Social Engineering | 5 | Financial Incentive | 5 | 50 | 60% | 20 | 80% | 24 |
| 5 | Reputation | CEO | 5 | Health & safety | 4 | failure of water/ power sup | 4 | 80 | 75% | 20 | 90% | 22 |
| 5 | Reputation | CEO | 5 | Health & safety | 4 | vironmental contamination | 4 | 80 | 75% | 20 | 90% | 22 |
| 5 | Reputation | CEO | 5 | Health & safety | 4 | Poor epidemic control | 4 | 80 | 75% | 20 | 90% | 22 |
| 1 | Staff/Temps | HR Manager | 2 | Death | 4 | Drug overdose | 5 | 40 | 60% | 16 | 80% | 19.2 |
| 1 | Staff/Temps | HR Manager | 2 | Death | 4 | Trip and Fall | 4 | 32 | 60% | 12.8 | 80% | 15.36 |
| 1 | Staff/Temps | HR Manager | 2 | Death | 4 | Traffic Accident | 4 | 32 | 60% | 12.8 | 80% | 15.36 |

**Planning**
Group 6 Healthcare Clinic involves the consultations of Asset Owners to determine the Asset values and then plan on how the case study is adopt and in line with ISO27001 standard, within the company framework. The stages will be assessing chronologically with management structure consulting IT Manager and his team and making them aware of the need to have risk assessment process and in compliance.

The IT systems plays a vital role in the company business continuity therefore an inventory will be conduct during this stage or assessment

**Analyzing**
All the data accumulated before are prepared to produce to generate the risk register. It is the obligation for the team to comply on achieving a better result for the company.

**Implementations**
There are three stages involve and are vital to adopt such measures to achieve good result, such as risk calculations, risk classification and risk control.
.
**Risk calculations**
When we create the risk register table, the following stage is to figure the potential risk on the recognized resources. Risk is ascertained by duplicating resource, risk and powerlessness esteems together. This is a quantitative approach in deciding the hazard related with every benefit.

**Risk Control**
Once chances are been ascertained and characterized, the following stage is to execute control to alleviate such hazard. In spite of the fact that, not all dangers are eradicate, but rather their effect can be decreased to a base. Risk control is a challenging factor of the risk management, because organization should try to control the risk at their level base on the company policy. Lack of expert within the organization to overcome the risk, organization tends to outsource their system and this very serious impact of risk control.

**Risks Classification**
Once the risk identify, the following stage is to group the hazard. Each association has their own technique in ordering dangers; the most common aspect is to construct them considering the greatness of their effect on the association. They can be delegate High, Medium or Low.

**Recommendations and Handover:**
After all risk has been assess and process the documents should be tender with recommendations for business continuation over to the Asset owner responsible.

**Maintenance (Training, Education, Updating and Monitoring):**

It is very important that after the gathering of the recorded risk register, the Asset owners are in communication with the case study team for business continuation and other recommendations that will further enable the company to adhere to its policy in compliance with the standard stipulated.

Additionally, the top managerial staff to endorse this recommendation once it has been acknowledged and the IT manager or a nominee from the company will make it official as an official standard to see that the company will enforce those recommendations.

**Result and Analysis:**
Risk Management has its own dimension in IT with cyber security being the main focal point due to cloud computing model being the main stay of organization especially government and healthcare organization. Every organization are always advice to secure risk register and develop Risk Appetite. Cultivate a risk culture, work out costing model, train staff, and explore new risk tools. This are built into an overall risk model of Risk Governance of IT and Cyber Security.

**Transferring of Risk**

This is when a risk control strategy attempts to transfer the risk to another entity. There are several ways to transfer the risk, considering the clinic after identifying the risk, we can use a best possible option for the transferring the risk by revising development models to achieve the goal of transfer or submitting the models to outsourcing to another company whose responsibilities are to mitigate risk. One final aspect is to contact the services provider, but in the case of the clinic, the software was in-house-built-in, so this option may not be applicable. The best practice is that every organization should stay within their product line to avoid conflict of interest. Giving an example like General Motors (MICHAEL WHITMAN & Herbert J. Mattord, Management of Information Security) who known for design and manufacturing cars and trucks. If they suggest expanding to other avenue of business type like energy and resources, they will tend to go for outsourcing, and will fail to invest to hire experts who will handle the information security aspect of their business. When such situation arises, it would be costly to the business or eventually might go down. Most organization fail in this situation when they should think of hiring expert to handle the Information security aspect or the organization to avoid such disaster to accrue. Failure to consider these factors, when situation arise, then the transfer of the risk will be outsourcing the risk or revising the development models. To transfer the risk is to implement a service level agreement between the two companies. The element involves are the services category, agreeing to services quality and measuring the services.

**Accepting of Risk**

This an approach is where the organization will take decision to do nothing if incident occur, but accept the results of the exploitation from the risk, unlike the mitigation approach to reduce the risk. In most case if the system compromise is about £20.000, the management will accept the risk control strategy to allow the system to go downhill and later replace it entirely.

**Avoidance of Risk**
This approach is where the organization will be implementing risk controls strategy mechanism (defense), in preventing exploitation of vulnerability. By considering the clinic it will put in place mechanism to avoid counter threats and quickly remove any vulnerabilities in its assets; these is done by implementing limiting access to assets like the server room and also add protection to safeguard the entire system. There are factors outline for any responsible organization to consider implementing in defense to safeguard their assets. Application of policy is one important aspect; this is where top management have the mandatory levels to certain produces to be follow by all. The control of passwords to IT System should be strictly assign and control. Staff training is eminent to educate staff on awareness of the dangers in the IT industry. If such training are constantly carry out by organizations, it might not to stop the risk, but will reduce the damage of the risk if accrue. Because, technology is day-to-day improving and technics are being develop day by day and counter measures are in the increase by cybercrimes. All these factors if implemented, will counter the threats facing an assets of the clinic.

**Mitigating of Risk**

Risk Mitigation is an approach of planning and preparing to reduce the damage caused by an incident, example for the clinic if problems rise, there should be a control mechanisms or contingency plan to reduce the risk before it affect the entire system. If any responsible organization wish to achieve its goal in combating this approach, they must consider three factors, the incidence response plan (IR), the disaster recovery (DR) and the business continuity plan, which will allow the business of the day to continue without interruption is consider as plan (B) in a business point of view.

**Reflections on the three Research Papers in the Moodle Handbook:**

The cloud-computing model also bought in a new wave called cyber security (Schatz et al 2017). The term cyber security is still vague in its definition but closely associate with the risk associated to IT information available of the internet (cyber). However, this can be Intranet, Internet or Extranet based. Most Government, HealthCare organization are now going into a B2G Business to government model of internet computing based on cloud infrastructure and this terminology are becoming more of a scope of clear definition. The EU General Data Protection Regulation (GDPR) is also just coming in terms with the definition of Cyber Security with its union and various governments. The term Cyber Security is not clearly defined but emphasized. In the past, action for breach break-in like "Talk-Talk cyber breach which occurred in the UK in October 2015" and was penalized by a £400,000 fine, in GDPR it can be up to 4% of the annual turnover of Talk-Talk, which is between EUR17m to EUR20m. James (2017). Repeated data breach penalized more as the organization failed to take care of the breach –though a case study has said it cannot be clearly state that there is a link directly between data breach and remediation done to the next attack (Schatz et al 2016).

The part focuses on examining the security breaches of information security and protection of information. Mainly due to a widespread global problem that information theft and breach is increasing day by day and becoming more damaging, coursing denial of services to authorize users.

The lack of sharing of data breaches by organization publicly is also a concern and hampers to tackle the problem, regardless the efforts to implement security controls to prevent information security breaches. As the problems continue to develop or not solved, organization still suffers in the hands of cybercriminals and a less guidance in literature on the term data breach.

A reach methodology describes the challenges of information security management, which has impact on the world economic due to security breaches and seen as cost but not moneymaking aspect. It also went ahead to investigates the impact on stock price announcement of organization asset value, which was widely seen as provocation of corporate control, regulatory policy and macroeconomic condition that affect the market value. Industrial reports have it to say that cost taking on various aspect, such as how costly detection outlays, escalation and notification of economic impact on organizational lost or diminished customer trust and confidence approach relaying on assumption of efficient rational markets result.

The best practice is to understand the role information security in a wider context and in an economist's point of view the challenging that are faced by organization, the best possible approach used by economists, the methodology applied to evaluated economic impact of information security events. By applying, an excellent mythology will yield to quality data. Management of Risk is a major issue to every organization, most organization fail to understand what risk is and how to map the risk. In the last decade, a model of risk management in IT has emerged with various tangible forms which make risk management an organization-wide activity led by a CRO along with CEO and IT managers.

IT based Risk is divided into many known elements like IT Infrastructure based risk which is server, network equipment's, desktop, servers. This risk is a straight-line risk and can be resolve by IT professional by putting measures such as installing the latest firewall and fine-tuning servers, desktops, anti-virus, anti-spyware, backup and having a recovery contingency policy in place.

The birth of mobile device added its own risk into the organization especially healthcare software. Various solutions emerged like windows azure, mobile iron, VMware air watch, spice work all have in-depth defense built on like data leak prevention policy, correlation attack, repudiations. Most of the risk are taken care by this software.

The invention of cloud computing with VMWARE, HYPER-V, Citrix software has saved a lot of infrastructure cost. However, it has emerged with a lot of its own security risk. Consolidating your server resource into one main host by virtualization and moving it around data Centre has helped business contingency, downtime, crash handling etc. The security risk of this has its own dimension. One of the main dimension being use of complex software now to handle the physical host, Virtual Machine (VM's) , SAN, Virtual Switches, Data Centre Switches, VCenter. The risk pattern has changed from low volatile single server to infra-based server which mismanaged can bring the whole organization down. New software like blue lane, safe has all entered the market foray.

**CONCLUSION:**

Cyber security threat has become a sophisticated and significant challenge that continues to grow. Organizations need the right outlines and management systems in order to come up with adequate measures that protect sensitive information. In order to attain this aim, organizations may have to consider developing more devolved cyber security risk management model that would enable them to come to a corporate understanding of the risk that relies on assessment and management by those who have ownership of the data. These systems should ensure that an organization:
- Is able to identify, evaluate and monitor cyber security risks
- Employ effective and secure data management practices and attitudes
- Implements and maintains appropriate network controls, this should include general controls and targeted security measures.

In order to achieve effective security, it is the responsibility of the whole organization and not only individual entrusted with the responsibility or asset owners. Effective security measures rely on an active collaboration between general staff, data/record collectors and all network users. In addition, it is important to keep active collaboration between other organizations, this helps to keep abreast of the rapidly evolving threats and enable the organization to prepare and protect themselves adequately.

Finally, achieving an effective and proportionate approach to managing cyber security, organizations such as The Healthcare Clinic will be able to maintain internal and external confidence to its staff, patients and the community as a whole, and continue to develop their core element of providing services safely and securely in the digital age.